\documentclass[osajnl,twocolumn,showpacs,superscriptaddress,10pt]{revtex4-1} 
\usepackage{amsmath,amssymb,graphicx}

\begin{document}


\title{Passive $\mathcal{PT}$-symmetry breaking transitions without exceptional points in dissipative photonic systems}
\author{Yogesh N. Joglekar}
\email{Corresponding author: yojoglek@iupui.edu\\© 2018 Optical Society of America]. Users may use, reuse, and build upon the article, or use the article for text or data mining, so long as such uses are for non-commercial purposes and appropriate attribution is maintained. All other rights are reserved.\\When adapting or otherwise creating a derivative version of an article published under OSAs OAPA, users must maintain attribution to the author(s) and the published article's title, journal citation, and DOI. Users should also indicate if changes were made and avoid any implication that the author or OSA endorses the use.}
\author{Andrew K. Harter}
\affiliation{Department of Physics, Indiana University-Purdue University Indianapolis (IUPUI), Indianapolis 46202, Indiana, USA}

\begin{abstract}
Over the past decade, parity-time ($\mathcal{PT}$)-symmetric Hamiltonians have been experimentally realized in classical, optical settings with balanced gain and loss, or in quantum systems with localized loss. In both realizations, the $\mathcal{PT}$-symmetry breaking transition occurs at the exceptional point of the non-Hermitian Hamiltonian, where its eigenvalues and the corresponding eigenvectors both coincide. Here, we show that in lossy systems, the $\mathcal{PT}$ transition is a phenomenon that broadly occurs without an attendant exceptional point, and is driven by the potential asymmetry between the neutral and the lossy regions. With experimentally realizable quantum models in mind, we investigate dimer and trimer waveguide configurations with one lossy waveguide. We validate the tight-binding model results by using the beam propagation method analysis. Our results pave a robust way toward studying the interplay between passive $\mathcal{PT}$ transitions and quantum effects in dissipative photonic configurations. 
\end{abstract}

\ocis{(080.1238) Array waveguide devices; (270.5585) Quantum information and processing.}
\maketitle 

\section{Introduction}
\label{sec:intro}
A fundamental principle of traditional quantum theory is that the observables of a system are Hermitian operators~\cite{qmbasics}. This self-adjoint character of observables is defined with respect to a global (Hamiltonian-independent) Dirac inner product. In particular, the Hamiltonian of a closed quantum system is Hermitian. It determines the energy levels of the system and therefore the experimentally observable transition frequencies. Thus, it came as a great surprise when Carl Bender and co-workers discovered a broad class of non-Hermitian, continuum Hamiltonians with purely real spectra~\cite{bender1,bender2}. The salient feature of such Hamiltonians is the presence of complex potentials $V(x)$ that are invariant under the combined parity ($\mathcal{P}:x\rightarrow -x$) and time-reversal ($\mathcal{T}=*$) operations, i.e. $V^*(-x)=V(x)$. Initial efforts on this subject focused on developing a self-consistent complex extension of quantum mechanics via a redefinition of the inner product that is used to define the adjoint of an observable~\cite{ajp,bender3}. This line of inquiry led to significant mathematical developments in understanding the properties pseudo-Hermitian operators~\cite{mostafa,mostafa1,mostafa2,mostafa3,mz1,mz2}. But it did not elucidate a simple physical picture for complex potentials that are a hallmark of $\mathcal{PT}$-symmetric Hamiltonians. 

Experimental progress on the $\mathcal{PT}$-symmetric systems started with two realizations~\cite{th1,th2}. First, Schrodinger equation is isomorphic with paraxial approximation to the Maxwell's equation where the local index of refraction $n(x)=n_R(x)+in_I(x)$ plays the role of the potential $V(x)$. Second, it is easy to engineer a lossy index of refraction $n_I(x)<0$, and not too difficult to engineer a gain either, i.e. $n_I(x)>0$. These dual realizations provided a transparent, physical insight into the meaning of complex, $\mathcal{PT}$-symmetric potentials: they represent balanced, spatially separated loss and gain~\cite{review}. Since then, over the past decade coupled photonic systems described by non-Hermitian, $\mathcal{PT}$-symmetric effective Hamiltonians have been extensively investigated~\cite{expt2,expt3,expt4,expt5,expt6,expt7,expt8,expt9}. Light propagation in such systems shows non-trivial functionalities, such as unidirectional invisibility~\cite{uni1,uni2}, that are absent in their no-gain, no-loss counterparts. We emphasize that these realizations are essentially classical. Gain at few-photons level is random due to spontaneous emission~\cite{gsa,blara}; in contrast, loss is linear down to single-photon level. Thus, engineering a truly quantum $\mathcal{PT}$-symmetric systems is fundamentally difficult. Therefore, in spite of a few theoretical proposals~\cite{2018review}, there are no experimental realizations of such systems that show quantum correlations present. 

By recognizing that a two-state loss-gain Hamiltonian is the same as a two-state loss-neutral Hamiltonian apart from an ``identity-shift" along the imaginary axis, the language of $\mathcal{PT}$-symmetry and $\mathcal{PT}$ transitions has been adopted to purely dissipative, classical systems as well~\cite{szameit1}. Indeed, the first ever observation of $\mathcal{PT}$-symmetry breaking was in two coupled waveguides, one with loss $\gamma$ and the other without~\cite{expt1}. As the loss strength was increased from zero, the net transmission first decreased from unity, reached a minimum, and then increased as $\gamma$ was increased beyond a threshold, signaling the passive $\mathcal{PT}$ symmetry breaking transition. This mapping provided a clear way forward to define passive $\mathcal{PT}$-symmetry breaking phenomenon in truly quantum, effectively two-level, dissipative systems. It led to the first observations of $\mathcal{PT}$-symmetry breaking in the quantum domain with quantum-correlated single  photons~\cite{pengxue} and ultracold atoms~\cite{leluo}. 

In this paper, we extend the notion of passive $\mathcal{PT}$ transition to lossy Hamiltonians that do not map onto a $\mathcal{PT}$-symmetric Hamiltonian. By using tight-binding models and beam propagation method (BPM) analysis of experimentally realistic setups, we demonstrate that such transitions -- driven by the emergence of a slowly decaying eigenmode -- occur without the presence of exceptional points. 

\section{Avoided level crossing in gain-loss systems}
\label{sec:alc}
The prototypical effective Hamiltonian for experimentally realized, classical gain-loss systems is given by $H_{PT}(\gamma)=-J \sigma_x+i\gamma \sigma_z\neq H^{\dagger}_{PT}$ where $\sigma_k$ are the standard Pauli matrices. This Hamiltonian is invariant under combined operations of $\mathcal{P}=\sigma_x$ and $\mathcal{T}=*$. With $\hbar=1$, the parameters of the Hamiltonian $H_{PT}$ have units of s$^{-1}$. In optical settings, the time is proportional to the distance $z$ traveled along the waveguide; therefore, the inter-site coupling $J$ is inversely proportional to the coupling length $L_c$, i.e. $J=\pi c/(n_0L_c)$ where $c/n_0$ is the constant speed of light in the waveguide with local index $n_0$. Similarly, the loss $\gamma$ and the potential offset $\delta$ (measured in s$^{-1}$) are linearly proportional to the inverse penetration depth and the propagation-constant offset (measured in m$^{-1}$) respectively. The eigenvalues of $H_{PT}$ are given by $\pm\sqrt{J^2-\gamma^2}$, are real for $\gamma\leq J$, and become complex conjugate pair for $\gamma>J$. The transition from the $\mathcal{PT}$-symmetric phase, i.e. purely real eigenvalues, to $\mathcal{PT}$-symmetry broken phase, i.e. some complex-conjugate eigenvalues, occurs at the threshold $\gamma_{PT}=J$. At this point, eigenvalues and eigenvectors of $H_{PT}(\gamma)$, both, become degenerate. Thus the $\mathcal{PT}$-symmetry breaking transition point is the same as the second-order exceptional point of the Hamiltonian $H_{PT}$~\cite{kato}. 

\begin{figure}[h]
\includegraphics[width=\columnwidth]{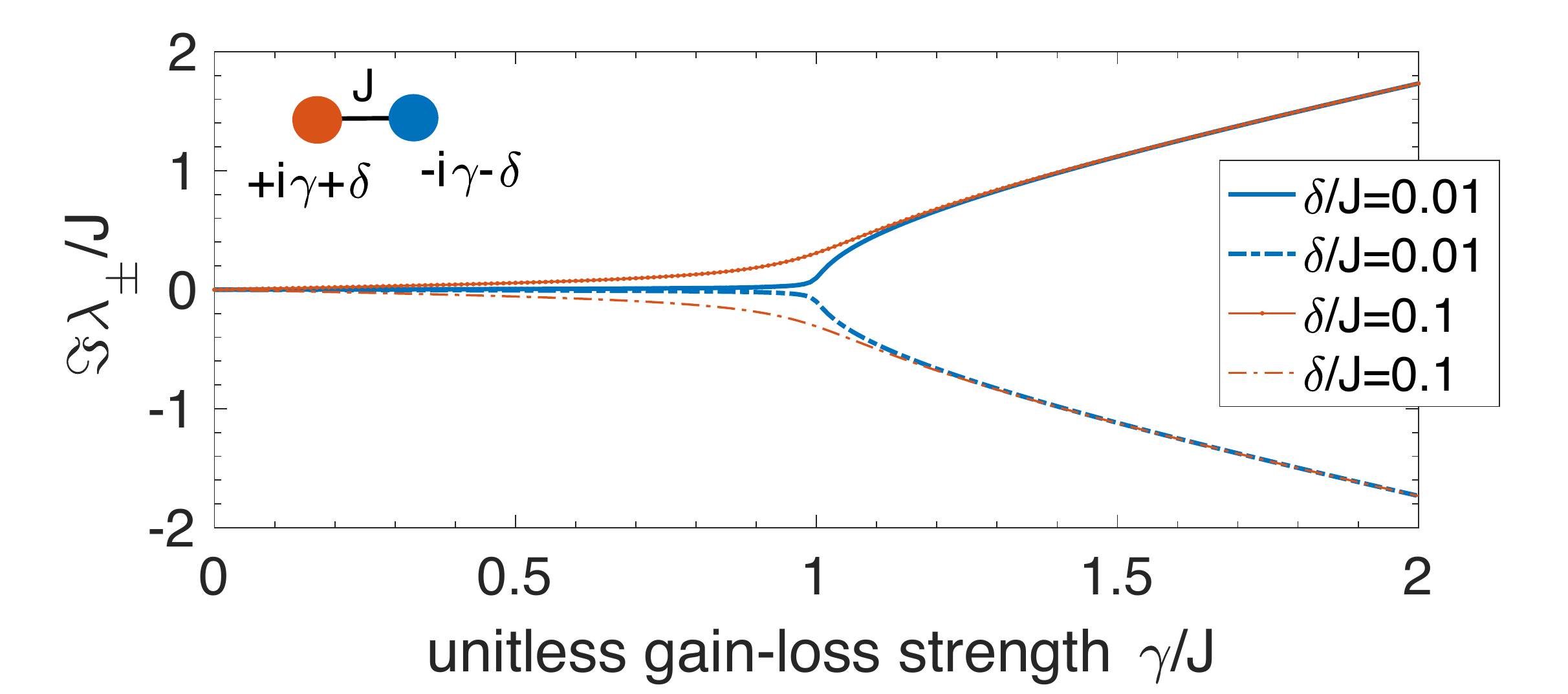}
\caption{Imaginary parts of the spectrum Eq.(\ref{eq:lambda2}) for nonzero perturbations $\delta/J=\{0.01,0.1\}$. The schematic $\mathcal{PT}$-dimer with gain (red) and loss (blue) sites is shown. When $\gamma/J\ll 1$, $\Im\lambda_{\pm}$ grow linearly with the gain-loss strength, but are nonzero. The divergence of their derivative at the threshold $\gamma=J$ is smoothed out as $\delta>0$ increases. In this case, the system is always in the $\mathcal{PT}$-broken phase, no matter how small $\Im\lambda_{\pm}(\gamma)\neq 0$ are.}
\label{fig:pt2flat}
\end{figure}

What happens when this Hamiltonian is perturbed by an antisymmetric real potential? Without loss of generality, such potential can be implemented by $i\gamma\rightarrow i\gamma+\delta$. The eigenvalues of the perturbed Hamiltonian $H(\gamma,\delta)$ immediately become complex,
\begin{equation}
\label{eq:lambda2}
\lambda_{\pm}(\gamma,\delta)=\pm\sqrt{ J^2-\gamma^2+\delta^2-2i\gamma\delta}.
\end{equation}
Figure~\ref{fig:pt2flat} shows that when $\delta>0$, for $\gamma\ll J$, the imaginary parts  $\Im\lambda_{\pm}$ are quite small and grow linearly with $\gamma$. This behavior changes to 
a steep $d\Im\lambda_{\pm}/d\gamma$ at $\gamma=J$, signaling the enhanced sensitivity in the neighborhood of the exceptional point~\cite{ep1,ep2}. However, the eigenvalues are always complex and therefore, the system is never in the $\mathcal{PT}$-symmetric phase when $\delta\neq 0$. Thus, the notion of $\mathcal{PT}$ symmetry breaking transition cannot be extended to perturbed gain-loss Hamiltonians $H(\gamma,\delta)$. 

In truly gain-loss systems, the $\mathcal{PT}$-symmetry breaking, defined by real-to-complex spectrum, occurs at the exceptional point, defined by coincidence of both eigenvalues and eigenvectors. Note that similar avoided level crossings~\cite{weissALC,rotterALC,rotterALC2} also occur in coupled lasers with static cavity losses and pump-current controlled gains, and show surprising phenomena such as pump-induced laser death~\cite{rotter2012,expt6}, loss-induced revival of lasing~\cite{expt7}, and laser self-termination~\cite{lstpra,lsttrimer}.


\section{PT-symmetry breaking in dissipative systems}
\label{sec:ptb}
The prototypical Hamiltonian for experimentally realized neutral-loss systems is given by 
\begin{equation}
\label{eq:hd}
H_{D}(\gamma)=-J\sigma_x-i\gamma(1-\sigma_z)=-J\sigma_x-2i\gamma|2\rangle\langle 2|
\end{equation}
In the quantum context, this represents a two-state system with loss only in the second state. In the classical context, this represents two evanescently coupled waveguides, where there is absorption only in the second waveguide~\cite{expt1}. Note that $H_{D}(\gamma)=H_{PT}(\gamma)-i\gamma{\bf 1}_2$; however due to the imaginary shift that is not invariant under time-reversal operation, the dissipative Hamiltonian does not commute with the antilinear $\mathcal{PT}$ operator, i.e. $[\mathcal{PT},H_D]\neq 0$. The eigenvalues of Eq.(\ref{eq:hd}) are given by $\lambda_{D\pm}=-i\gamma\pm\sqrt{J^2-\gamma^2}$ and always have positive decay rates $\Gamma_\pm=-\Im\lambda_{D\pm}$ for the two eigenmodes of the lossy Hamiltonian $H_D$. When $\gamma\leq J$, the two decay rates are equal and they increase linearly with $\gamma$, i.e. $d\Gamma_{\pm}/d\gamma>0$. Traditionally, this region is called the $\mathcal{PT}$-symmetric phase. When $\gamma=J$, the corresponding eigenmodes become degenerate. However, when $\gamma>J$, the decay rate for one of the eigenmodes starts to becomes smaller,
\begin{equation}
\label{eq:ld2} 
\Gamma_{+}=\gamma-\sqrt{\gamma^2-J^2}\xrightarrow[\gamma\gg J]{}\frac{ J^2}{2\gamma},
\end{equation}
while the second mode decays faster with $\Gamma_{-}\rightarrow 2\gamma$. Traditionally, this region is called the $\mathcal{PT}$-broken phase. We define the emergence of a slowly decaying mode, i.e. 
\begin{equation}
\label{eq:gpt}
\frac{d\Gamma}{d\gamma}=0 \hspace{2mm}\mathrm{at}\hspace{2mm} \gamma=\gamma_{PT},
\end{equation}
as the defining characteristic of the passive $\mathcal{PT}$ transition in loss-neutral systems~\cite{expt1,pengxue,leluo}. Indeed, the increased total transmission with increasing loss seen in Ref.~\cite{expt1} is due to this mode. We note that this criterion, defined by the sign of $d\Gamma/d\gamma$ changing from positive to negative for one of the modes, is not equivalent to defining the transition by the presence of an exceptional point, where the two decaying eigenmodes coalesce. For the identity-shifted Hamiltonian $H_D(\gamma)$, Eq.(\ref{eq:hd}), they happen to coincide. The latter criterion is only meaningful for the identity-shifted cases, whereas Eq.(\ref{eq:gpt}) is physically motivated and has straightforward experimental consequences~\cite{expt1,pengxue,leluo}. 

In the following sections we elucidate the consequences of this difference between dissipative $\mathcal{PT}$ systems and gain-loss systems for dimer and trimer models, while keeping evanescently coupled photonic waveguides in mind as their experimental realizations.

\section{Dissipative $\mathcal{PT}$ dimer and trimer cases}
\label{sec:disspt2}
Starting from the dissipative dimer Hamiltonian $H_D$, Eq.(\ref{eq:hd}), we now consider its Hermitian, on-site perturbation, 
\begin{equation}
\label{eq:disspt2}
H_2(\gamma,\delta)=\left(\begin{array}{cc}
0 & -J \\
-J & -2i\gamma-2\delta
\end{array}\right).
\end{equation}
Note that $H_2(\gamma,\delta)$ represents two waveguides with evanescent coupling $J$, with a loss potential $2\gamma$ in the second waveguide and an extra on-site potential $-2\delta$, that is generated by an offset in the real part of the index of refraction vis a vis the first waveguide. 

\begin{figure}[h]
\includegraphics[width=\columnwidth]{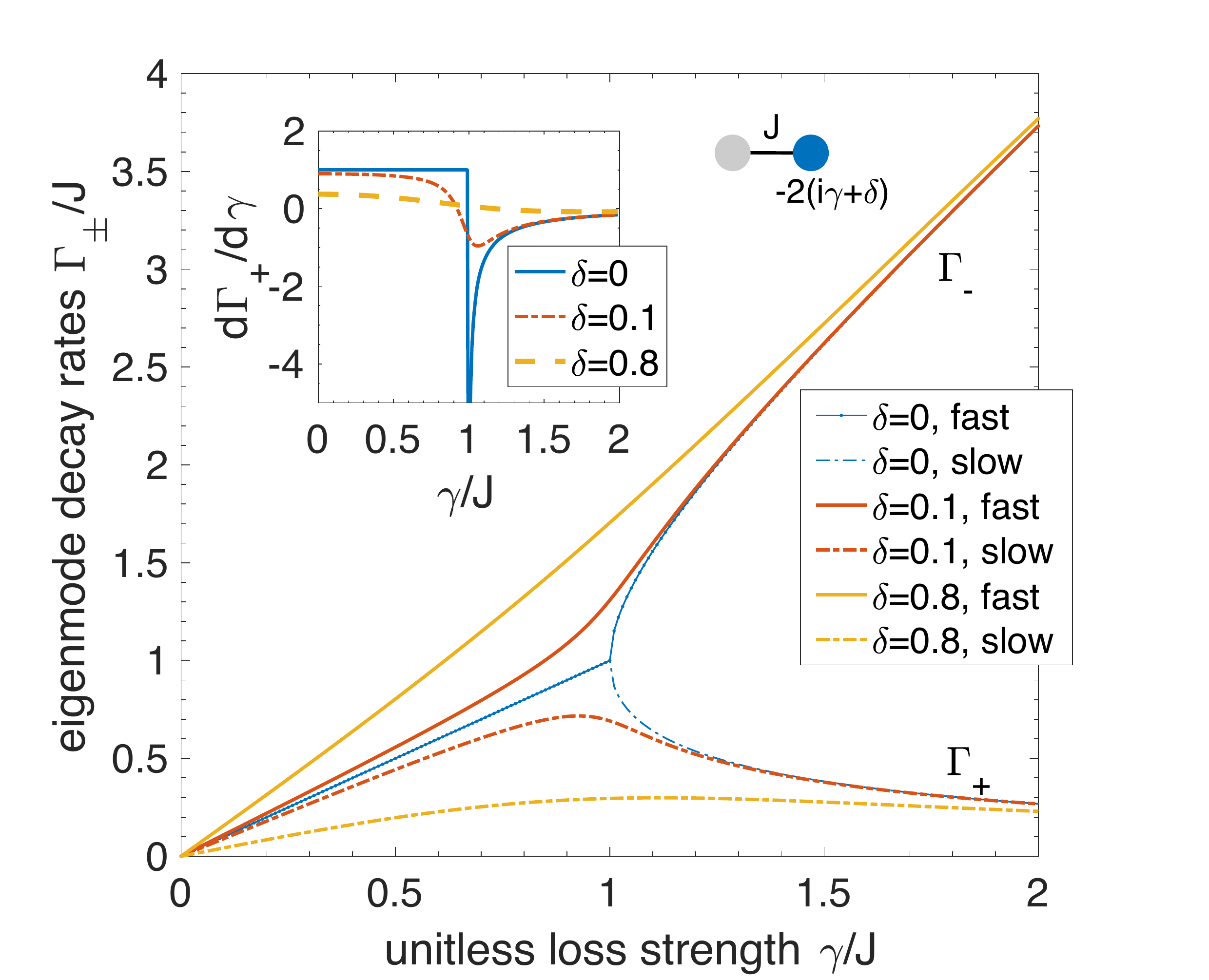}
\caption{Decay rates $\Gamma_{\pm}$ of the two eigenmodes of Eq.(\ref{eq:disspt2}) show the emergence of a slowly decaying mode for $\gamma\sim J$ not only at $\delta=0$, the prototypical exceptional point case, but also for a wide range of $\delta/J\neq 0$. Inset: the $\mathcal{PT}$ transition threshold is determined by $d\Gamma/d\gamma$ changing its sign from positive to negative. These results are even in the offset $\delta$ and thus remain the same for $\delta<0$.}
\label{fig:disspt2}
\end{figure}
It is straightforward to analytically obtain the eigenvalues $\lambda_{2D\pm}$ of $H_2(\gamma,\delta)$, similar to those in Eq.(\ref{eq:lambda2}), and the corresponding eigenvectors. Analyzing their behavior in the neighborhood of $\gamma/J=1, \delta=0$ requires further care, as it entails developing the corresponding Puiseux series in fractional powers of the distance from the exceptional point~\cite{kato,ep1,ep2}. In this paper, we primarily focus on numerical results across the entire parameter space instead of analytical results in the vicinity of the exceptional point. Figure~\ref{fig:disspt2} shows the evolution of the decay rates $\Gamma_\pm(\gamma)$ for the two eigenmodes as a function of the on-site perturbation. When $\delta=0$, the exceptional point and the passive $\mathcal{PT}$-symmetry breaking threshold, defined by the emergence of the slowly decaying mode, coincide. When $\delta/J=0.1$, the decay rates for the two modes are always different. Initially, both increase with $\gamma$ until near $\gamma\sim J$ where one takes off and the other starts to decrease, thus indicating the emergence of the slowly decaying mode. When the perturbation is increased further, $\delta/J=0.8$, a similar behavior is observed, but on a weaker scale. The inset in Fig.~\ref{fig:disspt2} shows the numerically obtained derivative of the decay-rate of the slow mode, $d\Gamma_{+}/d\gamma$. When $\delta=0$, we see that it is unity for $\gamma<J$ and diverges to $-\infty$ as $\gamma\rightarrow J^{+}$. For $\delta>0$, the inset provides a quick determination of the $\mathcal{PT}$ transition threshold $\gamma_{PT}$. These results are even in $\delta$, and the emergence of a slowly decaying mode is a robust feature for all offsets $\delta\neq 0$. We would like to point out that in the dissipative case the avoided-level-crossing location is instrumental to determining the passive $\mathcal{PT}$ transition point. In contrast, the counter-intuitive phenomena in coupled lasers do not occur at the avoided-level-crossing location, but at the location where an amplifying mode emerges~\cite{lstpra,lsttrimer}. 

\begin{figure}[h!]
\includegraphics[width=\columnwidth]{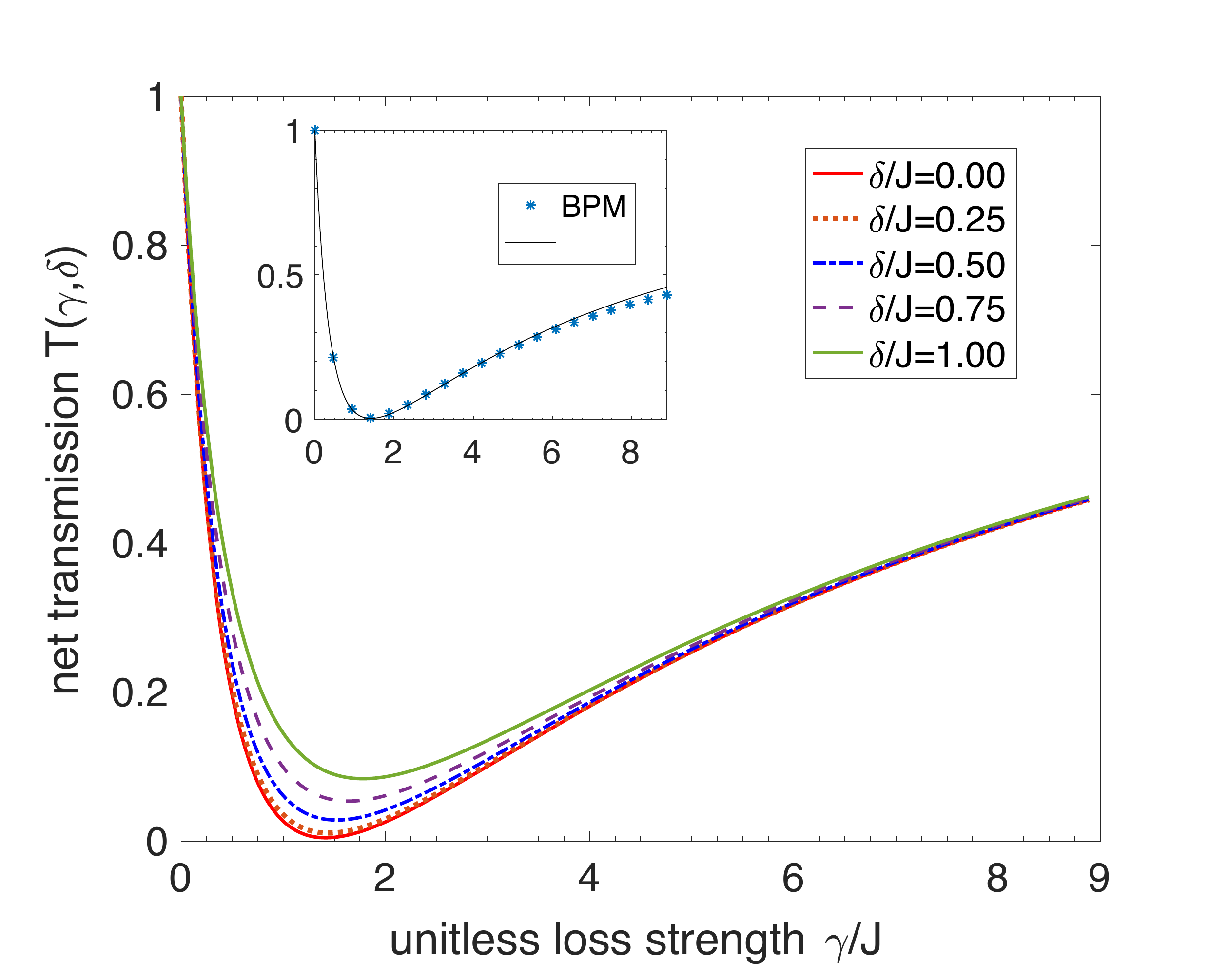}
\caption{Net transmission $T(\gamma)$ shows an upturn with increasing loss strength $\gamma$  signaling the passive $\mathcal{PT}$ transition~\cite{expt1}. It shows minimal change from its $\delta=0$ value~\cite{expt1} when $\delta/J$ is increased all the way to unity, i.e. the system is removed far from the exceptional point. These results are even in $\delta$ and thus remain unchanged for $\delta<0$. Inset: the lattice-model results (line) are consistent with the beam-propagation-method results (stars) obtained with sample parameters in Ref.~\cite{expt1}; see Sec.~\ref{sec:bpm} for details.}
\label{fig:transmit}
\end{figure}
There is no exceptional point for the Hamiltonian $H_2(\gamma,\delta\neq 0)$, and yet the passive $\mathcal{PT}$ transition phenomenon, defined by the emergence of a slow mode, is robust for $\delta\neq 0$. To show this, we obtain a key experimental signature~\cite{expt1}, the total transmission $T(\gamma,\delta)$ as a function of the on-site potential $\delta$. Starting from an initial injection into the first,  neutral site, i.e. $|\psi(0)\rangle=|1\rangle$, the total transmission at time $t$ (or equivalently at distance $z=ct/n_0$ traveled along the waveguide) is given by $T\equiv\langle\psi(t)|\psi(t)\rangle$ where $|\psi(t)\rangle=\exp(-iH_2t)|\psi(0)\rangle$. Keeping in mind Ref.~\cite{expt1}, we obtain the net transmission at a single-coupling length, i.e. $z=L_c$, or equivalently, $t=\pi/J$. Figure~\ref{fig:transmit} shows that $T(\gamma,\delta)$ has a robust upturn feature for all $\delta$, from $\delta=0$ to a large value of $\delta/J\sim 1$; it means that the key signature used in the passive $\mathcal{PT}$ symmetry breaking experiments~\cite{expt1} does not probe the exceptional point, but rather, the slowly decaying mode. The results in Fig.~\ref{fig:transmit} bolster the rationale for using Eq.(\ref{eq:gpt}) as the definition of passive $\mathcal{PT}$ symmetry breaking transition with or without exceptional points. 
 
\begin{figure}[h!]
\includegraphics[width=\columnwidth]{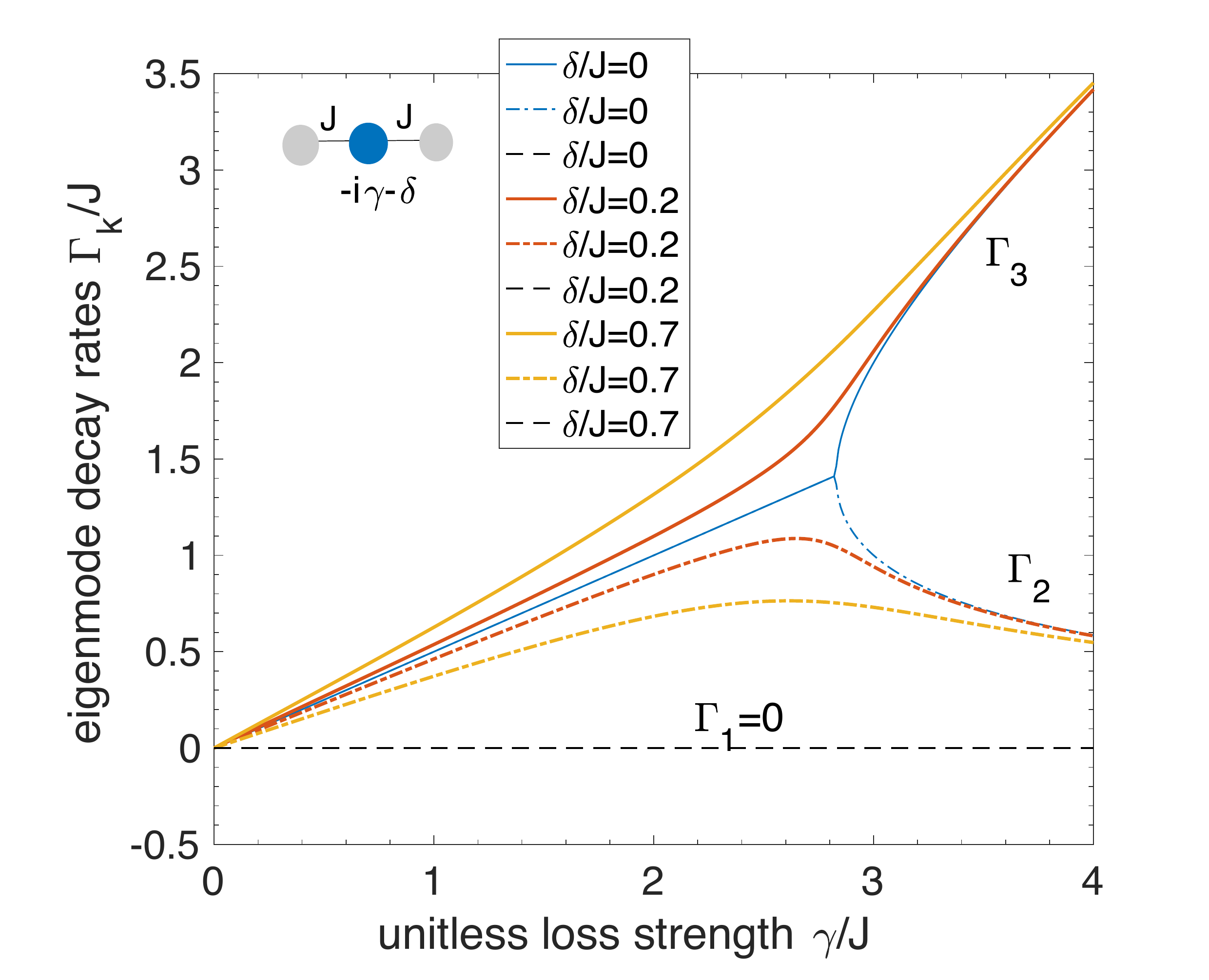}
\caption{$\mathcal{PT}$ transition in three coupled waveguides with lossy center waveguide, Eq.(\ref{eq:hc}). The decay rates $\Gamma_k$ show the emergence of a slow mode near $\gamma_{PT}\sim 2\sqrt{2}J$ for a wide range of center-site potential. These results are even in $\delta$ and thus remain valid for $\delta<0$.}
\label{fig:pt3hc}
\end{figure}

Let us now consider a dissipative trimer with only one lossy waveguide. The nearest-neighbor tunneling Hamiltonian for a trimer with open boundary conditions is given by $H_0=-J\left( |1\rangle\langle 2|+|2\rangle\langle 3| + \mathrm{h.c.}\right)$. With loss and a different on-site potential in the central waveguide, the trimer Hamiltonian becomes
\begin{equation}
\label{eq:hc}
H_{tc}=H_0-(i\gamma+\delta) |2\rangle\langle 2|. 
\end{equation}
Eq.(\ref{eq:hc}) has one zero eigenvalue with an antisymmetric eigenmode that does not couple to the center, lossy site. The other two eigenvalues are given by $[-(i\gamma+\delta)\pm\sqrt{8-(\gamma-i\delta)^2}]/2$. Figure~\ref{fig:pt3hc} shows the decay rates for the three modes as a function of the loss $\gamma$ and on-site potential $\delta$ in the central waveguide. The zero decay rate, $\Gamma_1=0$, denotes the antisymmetric eigenmode. When $\delta=0$, we see that the Hamiltonian $H_{tc}$ has an exceptional point at $\gamma_{PT}=2\sqrt{2}J$. For $\gamma<\gamma_{PT}$ both decay rates $\Gamma_2,\Gamma_3$ increase with loss strength, and the system is in the $\mathcal{PT}$-symmetric phase. It is characterized by a net transmission that decreases when $\gamma$ is increased. When $\gamma>\gamma_{PT}$, the decay rate $\Gamma_2$ starts to decrease and, as a result, the total transmission is increased with increasing $\gamma$~\cite{expt1,pengxue,leluo}. When $\delta\neq 0$ these features remain robust but the Hamiltonian $H_{tc}$ does not have an exceptional point. 

\begin{figure}[h!]
\includegraphics[width=\columnwidth]{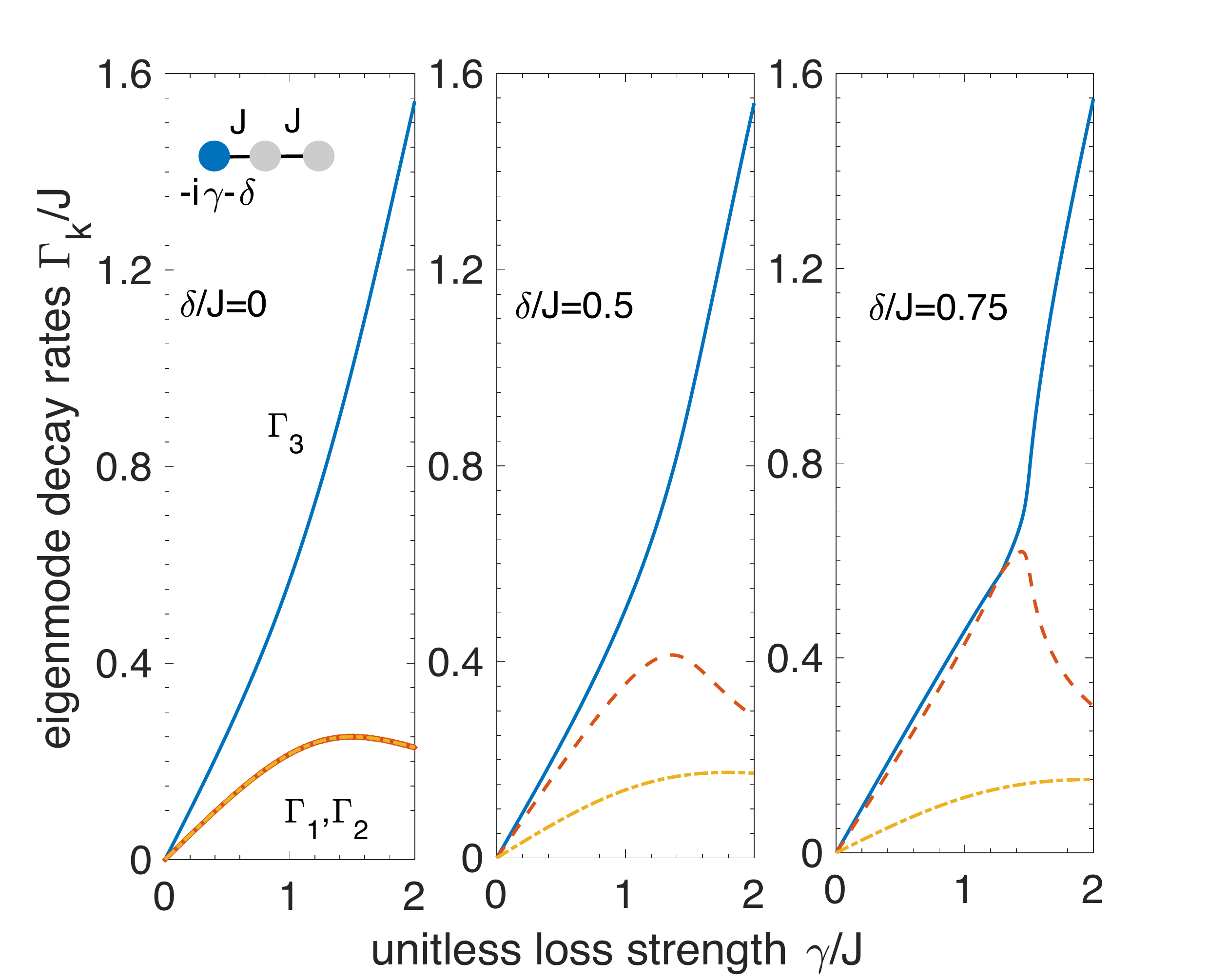}
\caption{$\mathcal{PT}$ transition in three coupled waveguides with lossy first waveguide, Eq.(\ref{eq:he}). When $\delta=0$, the two modes with equal decay rates undergo a sign-change for $d\Gamma_k/d\gamma$ near $\gamma/J\sim1.4$. When on-site potential $\delta>0$ is introduced, the two degenerate modes split, and the eigenmode decay-rate diagram hints at the existence of an exceptional point at $\delta/J=0.75$.}
\label{fig:pt3he}
\end{figure}

If the loss is in one of the outer waveguides, the trimer Hamiltonian is given by
\begin{equation}
\label{eq:he}
H_{te}=H_0-(i\gamma+\delta)|1\rangle\langle1|.
\end{equation}
Figure~\ref{fig:pt3he} shows the eigenmode decay rates $\Gamma_k$ as a function of the loss strength $\gamma$ for various on-site potentials $\delta$. When $\delta=0$, the trimer has two modes with equal decay rates, $\Gamma_1=\Gamma_2$, and a third one with the fastest decay rate $\Gamma_3$. Near $\gamma/J\sim 1.4$, the former two change over to slowly-decaying modes, signifying the $\mathcal{PT}$ transition. As $\delta>0$ is increased, the two degenerate decay rates split, and faster of the two, $\Gamma_2$, approaches the third one, $\Gamma_3$. Results in the last panel, corresponding to $\delta/J=0.75$, hint at the existence of an exceptional point in the vicinity of these parameters. In general, similar results are found for multiple losses and/or on-site potentials for the trimer model. The emergence of slowly decaying modes is commonplace, even when the existence of exceptional points is not. 

Lastly, we consider a trimer with periodic boundary conditions, with the Hamiltonian
\begin{equation}
\label{eq:hT}
H_{t}(\gamma,\delta,J')=H_{tc}-J'\left(|1\rangle\langle 3|+|3\rangle\langle 1|\right),
\end{equation}
where $0\leq J'\leq J$ denotes the tunable coupling between the first and the third waveguides. This ring configuration can be easily realized via femtosecond-laser direct-written waveguides in glass~\cite{szameit2}. We leave as an exercise for the reader to verify the following results. The antisymmetric mode continues to be an eigenmode of the Hamiltonian $H_t$ with real eigenvalue $+J'$. The other two decaying eigenmodes have eigenvalues given by 
\begin{equation}
\label{eq:lambdat}
\lambda_{t\pm}=-\frac{1}{2}\left(i\gamma+\delta+J'\right)\pm\frac{1}{2}\sqrt{8J^2+[(\delta-J')+i\gamma]^2}.
\end{equation}
It follows from Eq.(\ref{eq:lambdat}) that the exceptional point at $\delta=0$ shown in Figure~\ref{fig:pt3hc} survives along the line $\delta=J'$. When $\delta\neq J'$, it is also straightforward to check that a slow mode emerges at a loss strength $\gamma_{PT}\sim 2\sqrt{2}J$, even though the Hamiltonian $H_t$ does not have an exceptional point.  

It is worthwhile to mention that these trimer models cannot be identity-shifted to a $\mathcal{PT}$-symmetric model that supports a purely-real to complex-conjugate spectrum transition. That is, they are not isomorphic with any balanced gain-loss trimer models. They stand on their own and display the passive $\mathcal{PT}$ transition without an exceptional point over a wide range of Hamiltonian parameters. In the following section, we verify these results via beam propagation method (BPM) analysis of three planar coupled waveguides. 

\begin{figure*}
\includegraphics[width=\columnwidth]{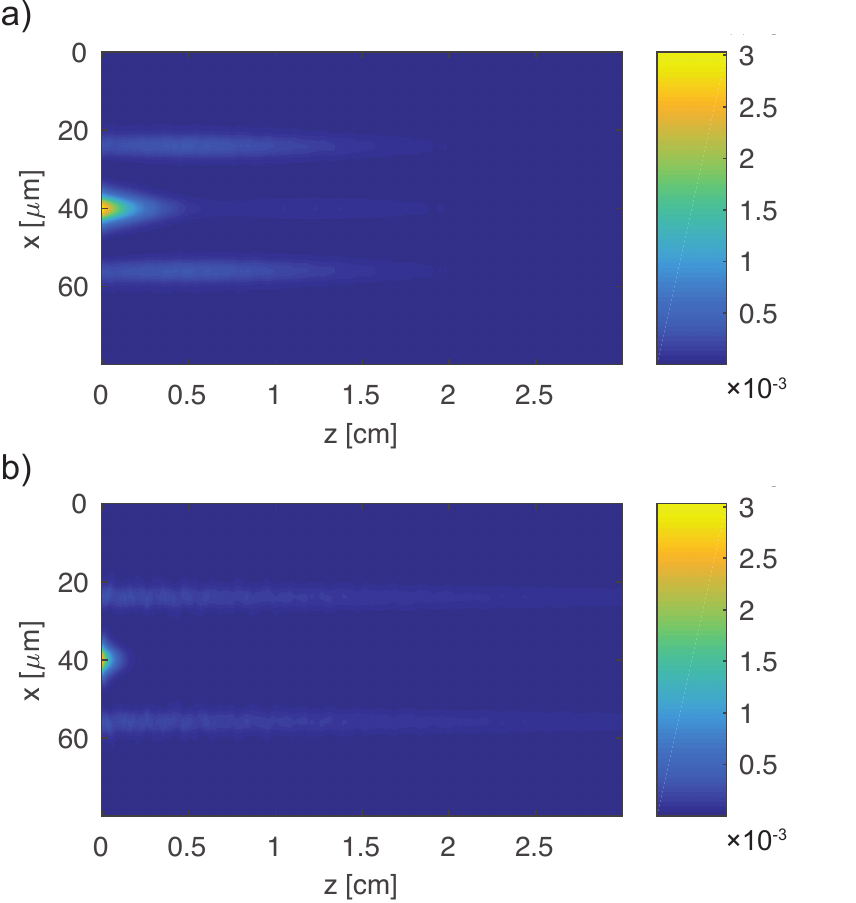}
\includegraphics[width=\columnwidth]{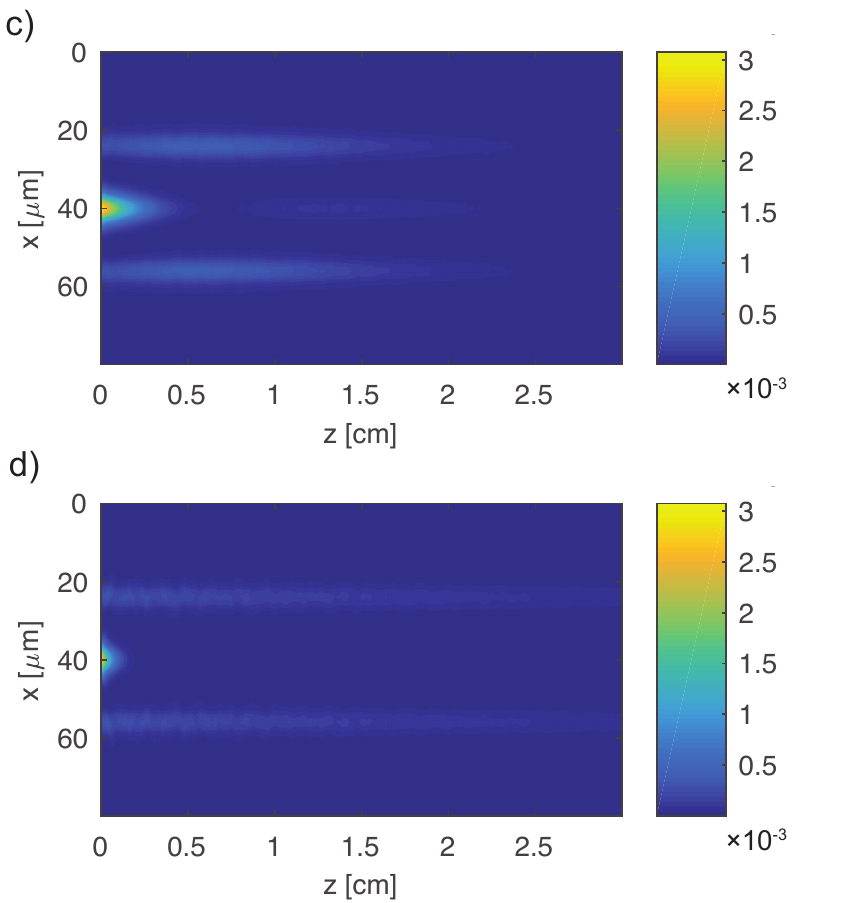}
\caption{BPM results for the intensity $I(x,z)$ of a symmetric initial pulse in a waveguide trimer defined by Hamiltonian Eq.(\ref{eq:hc}) with $\delta=0$ (a,b) and $\delta/J=1$ (c,d). (a) when $\gamma/J=2.75<\gamma_{PT}$, the pulse decays quickly. (b) at a much larger loss, $\gamma/J=10$, the pulse propagates longer, giving rise to increased transmission. (c) for $\gamma/J=2.75$, the pulse decays quickly. (d) at $\gamma/J=10$, the pulse propagates longer. In both cases, $\delta=0$ and $\delta/J=1$, the emergence of a slowly decaying mode is clear; in the latter case, the Hamiltonian is far removed from an exceptional point.}
\label{fig:bpm1}
\end{figure*}
\section{Beam propagation method analysis}
\label{sec:bpm}
The results presented in Sec.~\ref{sec:disspt2} are based on a tight-binding Hamiltonian where the spatial extent of a ``site" is ignored. In a coupled-waveguides realizations of a lattice model, the electric field strength across the waveguide is not constant, and its spatial variation in the direction transverse to the waveguide needs to be taken into account. We verify our tight-binding results by obtaining the time-evolution of the wavefunction $\psi(x,t)$ in a planar waveguide trimer with realistic parameters~\cite{zeuner}. We remind the reader that $\psi(x,t)$ is the envelope of the actual electric field, i.e., $E(x,t)=\exp\left[ik_0z-ic(k_0/n_0)t\right]\psi(x,t)$ where $k_0$ is the wavenumber of the rapidly varying part of the electric field, $c$ is the speed of light, and $n_0$ is the real index of refraction of the cladding. Under paraxial approximation, the envelope $\psi(x,t)$ obeys the continuum Schrodinger equation for a particle with an effective mass $m=k_0 n_0^2/c$ in a potential $V(x)=ck_0\left[1-n(x)^2/n_0^2\right]$. The position-dependent index of refraction $n(x)$ differs from that of a cladding only within each waveguide. For a small refractive-index contrast $\Delta n/n_0\sim 10^{-4}\ll 1$, the potential becomes $V_p=2ck_0\Delta n_p/n_0$ within waveguide $p$. We model the losses by adding negative imaginary parts to the index contrast $\Delta n_p$ and generate different on-site potentials by choosing different real parts while satisfying the small-contrast constraint. (For more details regarding these calculations, see Refs.~\cite{zeuner,aah,sr}.) 

Figure~\ref{fig:bpm1} a,b show the resultant intensity plots $I(x,z)=|\psi(x,z=ct/n_0)|^2 dx$ for a dissipative trimer with an exceptional point, $\delta=0$. The initial state is a symmetric combination of three single-waveguide eigenmodes with weight ratios 1:9:1. The symmetric combination ensures that the initial state is decoupled from the antisymmetric eigenmode with zero decay rate. Figure~\ref{fig:bpm1}a shows the rapid decay of the initial pulse when the loss strength $\gamma/J=2.75$ is just below the threshold $\gamma_{PT}=2\sqrt{2}J$. When the loss strength is increased four-fold to $\gamma/J=10$, Fig.~\ref{fig:bpm1}b, the same pulse travels farther signaling the emergence of the slowly decaying mode in the $\mathcal{PT}$ broken regime. Figure~\ref{fig:bpm1} c,d show corresponding results for a trimer with no exceptional point and a very large on-site potential $\delta/J=1$. We use such a large offset because the BPM simulations are virtually indistinguishable from the $\delta=0$ case for $\delta\lesssim J/2$. The eigenvalue flow in Fig.~\ref{fig:pt3hc} shows that in this case, after an initial linear increase, the decay rate of the slow-mode $\Gamma_2$ does not change appreciably at large $\gamma\geq\gamma_{PT}$. Figure~\ref{fig:bpm1}c shows that when $\gamma/J=2.75$ the initial pulse has a partial oscillation back into the central, lossy waveguide. When the loss strength is increased to $\gamma/J=10$, Fig.~\ref{fig:bpm1}d, the pulse travels farther indicating a $\mathcal{PT}$-broken regime. These representative BPM results are consistent with tight-binding model findings, and show that in dissipative systems, a $\mathcal{PT}$ transition can occur with or without exceptional points. 

\section{Discussion}
In this work, we have proposed a meaningful extension of the passive $\mathcal{PT}$ symmetry breaking phenomenon to dissipative Hamiltonians that are not identity-shifted from a balanced gain-loss Hamiltonian. We have shown that passive $\mathcal{PT}$ transitions, defined by the emergence of a slowly decaying mode, are ubiquitous in dissipative photonic systems and are not contingent on the existence of exceptional points. This is in sharp contrast with classical systems with balanced gain and loss, where $\mathcal{PT}$-symmetry breaking and exceptional points go hand in hand, and the concept of $\mathcal{PT}$ symmetry breaking cannot be meaningfully extended to unbalanced Hamiltonians. Our results may also be applicable to dissipative metamaterial systems, where non-trivial phenomena occur in the presence of balanced gain and loss~\cite{fu1,fu2}. 

Due to their exceptional tunability, relatively easy scalability, and the ability to implement one and (post-selected) two-qubit operations, integrated photonic systems at a quantum level are of great interest for quantum simulations and quantum computing. By introducing mode-dependent, tunable losses, such systems can be further tailored for true quantum simulations of non-Hermitian Hamiltonians in general and (dissipative) $\mathcal{PT}$-symmetric Hamiltonians in particular. Such simulators will permit the investigation of quantum attributes, such as entropy, entanglement metrics, many-particle correlations, etc. across the passive $\mathcal{PT}$ transition. Carrying out such studies in balanced gain-loss systems remains an open question and, most likely, is fundamentally impossible~\cite{szameit2018}. 

\section*{Acknowledgment} 
This work was supported by NSF grant no. DMR-1054020.


\end{document}